\documentclass[12pt]{iopart}
\usepackage{graphicx}
\usepackage{color}

\begin{document}

\title{A general group theoretical method to unfold band structures and its application}

\author{Huaqing Huang,$^1$ Fawei Zheng,$^{2,3}$ Ping Zhang,$^{2,3}$ \footnote{zhang\_ping@iapcm.ac.cn} Jian Wu,$^{1}$
Bing-Lin Gu,$^{4,5}$ and Wenhui Duan$^{1,4,5}$\footnote{dwh@phys.tsinghua.edu.cn}}
\address{$^1$Department of Physics and State Key Laboratory of Low-Dimensional Quantum Physics, Tsinghua University, Beijing 100084, People's Republic of China}
\address{$^2$LCP, Institute of Applied Physics and Computational Mathematics, Beijing 100088, People's Republic of China}
\address{$^3$Beijing Computational Science Research Center, Beijing 100084, People's Republic of China}
\address{$^4$Institute for Advanced Study, Tsinghua University, Beijing 100084, People's Republic of China}
\address{$^5$Collaborative Innovation Center of Quantum Matter, Tsinghua University, Beijing 100084, People's Republic of China}


\begin{abstract}
We present a general method to unfold energy bands of supercell calculations to primitive Brillouin zone using group theoretical techniques, where an isomorphic factor group is introduced to connect the primitive translation group with the supercell translation group via a direct product. Originating from the translation group symmetry, our method gives an uniform description of unfolding approaches based on various basis sets, and therefore, should be easy to implement in both tight-binding model and existing \textit{ab initio} code packages using different basis sets. This makes the method applicable to a variety of problems involving the use of supercells, such as defects, disorder, and interfacial reconstructions. As a realistic example, we calculate electronic properties of an monolayer FeSe on SrTiO$_3$ in checkerboard and collinear antiferromagnetic spin configurations, illustrating the potential of our method.
\end{abstract}

\pacs{71.15.-m, 71.20.-b, 73.20.-r}


\maketitle

\section{Introduction}

The electronic energy band structure (EBS) is a basic concept in textbooks of condensed matter physics \cite{solid} and can be used to investigate various physical properties of crystal materials. Furthermore, the EBS can be directly compared with the results of angle resolved photoemission spectroscopy (ARPES) measurement, which give the spectral function within the quasiparticle picture. Because crystal materials have translational symmetry, we can introduce wave vector $\vec{k}$ based on Bloch's theorem and get the energy $E(\vec{k})$ as a function of wave vector $\vec{k}$ (i.e., the EBS). With the development of computational techniques and the progress in condensed matter theory, one can easily calculate the EBS of perfect crystal materials using either empirical tight binding (TB) model \cite{slater} or first-principles methods within the framework of density functional theory (DFT). \cite{HK,KS}

However, when the translational symmetry of the physical systems is destroyed by defects (such as substitutional doping, impurity, vacancy, dislocation, \emph{etc}.), \cite{Heumen,TM,Edoping,Konbu} disorder, \cite{zunger_prl,e0g,haverkort} interfacial reconstructoin \cite{Kim,YQi,silicene} and even different spin configurations, \cite{KaiLiu} one cannot define $\vec{k}$ as well as EBS in the first Brillouin zone (FBZ) of the primitive lattice. It is common practice to use the supercell approximation with periodic boundary conditions in the computational exploration of those aperiodic systems. As the supercell becomes larger, the corresponding FBZ shrinks and the calculated EBS becomes gradually dense energy levels. Consequently, it is hard to extract useful information from such heavily folded EBS and compare it with ARPES results. In recent years, much effort has been devoted to resolve this issue. Boykin \textit{el. al.} developed basic ideas of unfolding
under tight binding approximation. \cite{boykin_prb71,boykin_prb76,boykin_jpcm} Ku \textit{et al.} \cite{weiku} developed an effective algorithm to unfold EBS of supercell to the FBZ of the primitive lattice via localized Wannier functions. Popescu \textit{et al.} \cite{zunger_prb} also presented a method to extract an effective band structure from supercell calculations on random alloys. Allen \textit{et al.} \cite{pba} provided a convenient notation and useful theoretical formulas of band unfolding which were applied in electron, phonon and slab systems.

Generally the supercell approximation introduces an artificial translational symmetry of the supercell lattice. Since the translational symmetry operations of primitive and supercell lattices form two Abian groups, we can deal with the unfolding process from the aspect of group theory. In this work,  by investigating the relationship between the above two translational groups, we propose an approach to unfold EBS of supercell calculation to primitive Brillouin zone. {This approach describes unfolding procedure using different basis sets uniformly, and thus is easy to implement in all TB models and \textit{ab initio} code packages where different basis sets such as plane waves, atomic orbtials or Wannier functions are employed.} We then demonstrate the validity of this method in a TB calculation of $30\times30$ graphene supercell. As a more realistic example, we further investigate the effect of magnetic order of monolayer FeSe on SrTiO$_3$ substrate, and find that checkerboard antiferromagnetic (AFM) instead of collinear AFM order might be the ground state spin configuration which yields the EBS similar to recent experimental observations. \cite{NComm,SLHe,DLFeng}

\section{Methodology}
\subsection{Translation group $G_p$, $G_s$ and $G_f$}
The periodic materials, like crystals, have translational symmetry. The translation operation $T_{\vec{r}}$ leaves the lattice invariant.  All the translation operations for primitive lattice form an infinite Abelian group, named as $G_p$. Besides the primitive lattice, supercell lattice is also frequently used in electronic structure calculations, which is suitable to simulate complicated systems such as defects and alloys. The translation operations $T_{\vec{R}}$, which leave the supercell lattice invariant, form another group, $G_s$. The translation vectors for group $G_p$ and $G_s$ can be written as
\begin{eqnarray}
  \vec{r}&=&\sum_{i=1}^D n_i\vec{a}_i, \\
  \vec{R}&=&\sum_{i=1}^D N_i\vec{A}_i,
\end{eqnarray}
where $n_i$ and $N_i$ are integers, and $D$ stands for the dimension of the system. The supercell basis vectors $\vec{A}_i$ and primitive basis vectors $\vec{a}_j$ are related by an integer matrix $N$, specifically, $\vec{A}_i=\sum^D_j N_{ij}\vec{a}_{j}$.

Actually, all the elements of group $G_s$ are also elements of group $G_p$, and thus, $G_s$ is a subgroup of $G_p$. Because both $G_s$ and $G_p$ are Abelian groups, there should exist a group isomorphic to the factor group (i.e., $G_p/G_s \cong G_f$), which is denoted by $G_f$ for simplicity. $G_p$ is the direct product of $G_s$ and $G_f$,
\begin{equation}
G_p=G_s\otimes G_f.
\label{Eq3}
\end{equation}
The translation vector of group $G_f$ has the same form of $\vec{r}$:
\begin{equation}
\vec{\mathcal{R}}=\sum^D_in_i\vec{a}_i.
\end{equation}
Different from infinite $G_p$ and $G_s$, $G_f$ is a finite group with the translation operations obeying
\begin{equation}
T_{\vec{\mathcal{R}}}=E ~~~~ (\mathrm{for~~} \vec{\mathcal{R}} = \vec{A}_i),
\label{Eq17}
\end{equation}
where $E$ is the unit element of $G_f$.

\subsection{Group representation and energy bands unfolding}
For Abelian groups, each group element is a class by itself, and then the dimension of irreducible unitary matrix representation is $1\times 1$ (just a complex number with norm 1). Thus, the irreducible representations for $G_p$, $G_s$ and $G_f$ can be written as
\begin{eqnarray}
D_{\vec{k}}(T_{\vec{r}})&=&e^{i\vec{k}\cdot\vec{r}} ~~~~~~ (\vec{k} \in \omega),\\
D_{\vec{K}}(T_{\vec{R}})&=&e^{i\vec{K}\cdot\vec{R}} ~~~~ (\vec{K} \in \Omega),\\
D_{\vec{\mathcal{K}}}(T_{\vec{\mathcal{R}}})&=&e^{i\vec{\mathcal{K}}\cdot \vec{\mathcal{R}}} ~~~~ (\vec{\mathcal{K}} \in \Lambda),
\end{eqnarray}
where $\vec{k}$ is in the FBZ of primitive lattice ($\omega$), and $\vec{K}$ is in the FBZ of supercell lattice ($\Omega$).
From Eq.~(\ref{Eq17}), the vector $\vec{\mathcal{K}}$ can be written as  $\vec{\mathcal{K}}=\sum^D_i \mathcal{N}_i \vec{B}_i$, where $\vec{B}_i$ are reciprocal lattice vectors of the supercell lattice. Herein $\mathcal{N}_i$ is an integer and should be chosen to ensure that $\vec{\mathcal{K}}$ is in the primitive cell FBZ $\omega$. The total number of choices for $\{\mathcal{N}_1,\cdots,\mathcal{N}_D\}$ is equal to the volume ratio $|\omega|/|\Omega|$, which is also the order of group $G_f$. We denote the set of all the possible $\vec{\mathcal{K}}$ by $\Lambda$.

According to Bloch's theorem, the basis function of the irreducible representation of $G_p$ can be written as
\begin{equation}
\psi_{\vec{k}}(\vec{x})=e^{i\vec{k}\cdot \vec{x}}u_{\vec{k}}(\vec{x}),
\end{equation}
where $u_{\vec{k}}(\vec{x})$ is a periodic function in the primitive lattice (i.e., $T_{\vec{r}}u_{\vec{k}}(\vec{x})=u_{\vec{k}}(\vec{x})$). As a result of the fact that $G_s$ is an subgroup of $G_p$, the function $\psi_{\vec{k}}(\vec{x})$ is also the basis function of group $G_s$,
\begin{equation}
\psi_{\vec{k}}(\vec{x})= e^{i(\vec{K}+\vec{\mathcal{K}})\cdot \vec{x}}u_{\vec{K}+\vec{\mathcal{K}}}(\vec{x})
= e^{i\vec{K}\cdot \vec{x}}[e^{i\vec{\mathcal{K}}\cdot \vec{x}}u_{\vec{K}+\vec{\mathcal{K}}}(\vec{x})]
= e^{i\vec{K}\cdot \vec{x}}u'_{\vec{K}}(\vec{x}).
\end{equation}
Herein the function $u'_{\vec{K}}(\vec{x})$ is periodic in the supercell lattice [$T_{\vec{R}}u'_{\vec{K}}(\vec{x})=u'_{\vec{K}}(\vec{x})$], and it is also the basis function of group $G_f$ with wave vector $\vec{\mathcal{K}}$.

In supercell electronic structure calculations, we get a series of functions $u'_{\vec{K}}(\vec{x})$ and eigenenergies for each $\vec{K}$. As discussed above, there is a hidden parameter $\vec{\mathcal{K}}$ in $u'_{\vec{K}}(\vec{x})$. Once we get the parameter $\vec{\mathcal{K}}$, the Bloch function and the corresponding energy for vector $\vec{k}=\vec{K}+\vec{\mathcal{K}}$ in $\omega$ are known. In other words, the energy bands are unfolded. {Since any primitive wavevector $\vec{k}$ in primitive FBZ belongs to a unique $\vec{K}$ in the supercell FBZ, an easy search in supercell reciprocal lattices can yield the proper $\vec{\mathcal{K}}$, so that $\vec{k}-\vec{\mathcal{K}}$ is in the supercell FBZ. More detailed analysis about wavevectors can be found in Refs. \cite{eurjphys,physE}}

In the following, we will present a method to identify the wave vector $\vec{\mathcal{K}}$ for $u'_{\vec{K}}(\vec{x})$. We choose a set of normalized orthogonal periodic functions $v_i(\vec{x})$ ($i=$1, 2, 3 ...), which are complete in the primitive cell (for the nonorthogonal case,
it is straightforward to perform the standard orthonormalization procedure priorly). The basis function of group $G_f$ can be written as $|i, \vec{\mathcal{K}}^\prime\rangle=e^{i\vec{\mathcal{K}}^\prime \cdot \vec{r}}v_i(\vec{x})$. Then we construct a group of projection operators $P(\vec{\mathcal{K}}^\prime)$ as
\begin{equation}
P(\vec{\mathcal{K}}^\prime)=\sum_i |i, \vec{\mathcal{K}}^\prime\rangle \langle i, \vec{\mathcal{K}}^\prime|  ~~~~ (\vec{\mathcal{K}}^\prime \in \Lambda).
\end{equation}
The expectation value of $P(\vec{\mathcal{K}}^\prime)$ for function $u'_{\vec{K}}(\vec{x})=e^{i\vec{\mathcal{K}}\cdot \vec{x}}u_{\vec{K}+\vec{\mathcal{K}}}(\vec{x})$ is $\delta_{\vec{\mathcal{K}}\vec{\mathcal{K}}^\prime}$. From this
expectation value, we can identify the vector $\vec{\mathcal{K}}$, then further unfold the energy bands. The unfolded energy bands should be identical to the energy bands calculated by using the
primitive cell for a perfect lattice. However, when the translational symmetry is broken (due to impurity, structure reconstruction, magnetic reconfiguration, \emph{etc}.), the
above unfolding procedure produces new energy bands, which are similar to (but need not be identical to) energy bands calculated by using the primitive cell. The expectation value of $P(\vec{\mathcal{K}}^\prime)$ for $u'_{\vec{K}}(\vec{x})$ is  between 0 and 1 instead of 0 or 1, representing the weight of $u'_{\vec{K}}(\vec{x})$ at point $\vec{K}+\vec{\mathcal{K}}^\prime$. Thus, the energy bands become fuzzy and may break into separate parts. In this case, the unfolded energy bands contain information of the defects or reconstruction and can be directly compared with ARPES or other measurements. {Since the main difference of computing real and complex bands lies in the choice of basis, our method can be used for complex bands too, if some modifications are included.\cite{ajoy2,ajoy}}

In practice, the function $v_i(\vec{x})$ can be taken as plane waves, atomic orbitals, Wannier functions or any other basis sets in quantum chemistry. Thus our method should be easy to implement in both TB models \cite{boykin_prb71,boykin_prb76,boykin_jpcm} and existing \textit{ab initio} code packages employing different basis sets. \cite{vasp,qe,siesta,nmto} Note that when we choose $e^{i\vec{\mathcal{K}}^\prime \cdot \vec{x}}v_i(\vec{x})$ to be the eigenvectors of
the primitive cell calculations, they are complete and can be represented by a plane-wave expansion, then we can get similar weight expression to Ref.~\cite{zunger_prb} {(see Appendix)}. In fact, the basis sets in our method should be complete periodic functions but are not required to be eigenstates. It turns out that the method is even simpler when the local basis sets are used. For example, when Wannier functions of the primitive cell are used as the $v_i(\vec{x})$, the method turns out similar to Ku \textit{et al.}'s approach. \cite{weiku} Other localized basis sets can also be adapted, which result in a modified formula of weight factor. \cite{CCLee} Hence, our method provides a more general strategy to unfold EBSs by constructing a group of universal projector operators which are not restricted to specific basis sets.

\section{Numerical implementation and applications}\label{sec:appl}

In this section we discuss the application of the above-described method. First, we illustrate the validity of the approach in $30\times30$ graphene supercell calculations within the TB approximation. We then apply the method to investigate the magnetic order of monolayer FeSe on TiO$_2$-terminated SrTiO$_3$ (001) surface, which shows signatures of high temperature superconductivity with $T_c > 50$ K, \cite{QYWang} within the framework of DFT.

\subsection{Unfolding bands of perfect graphene in the TB calculations}

\begin{figure}[bc]
\begin{centering}
  \includegraphics[width =0.5\textwidth]{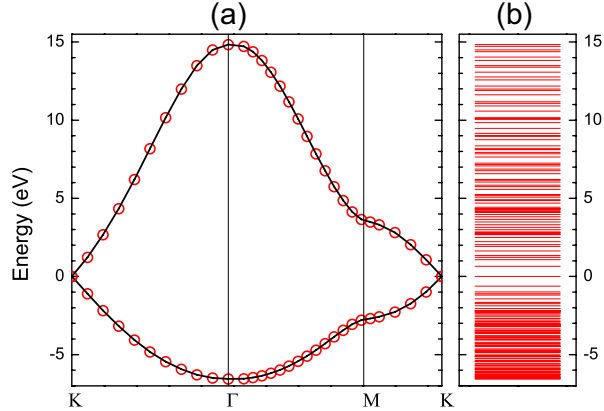}\\
  \caption{The TB energy bands of graphene from primitive cell calculation
[black line in panel (a)] and supercell calculations with band unfolding [red circles in panel (a)].
The unfolded energy bands are obtained from the single $\Gamma$ point calculation of
$30\times30$ supercell [panel (b)].
}\label{1}
\end{centering}
\end{figure}

Graphene is a one atom thick two dimensional allotrope of carbon with unique Dirac cone in its EBS, which can provide a simple but nontrivial test for our unfolding method. We consider the $30\times 30$ supercell for graphene in the TB calculation. Herein, only one $p_z$ orbital for each carbon atom is used, and the TB parameters are taken as transfer integral $t=-3.03$ eV and overlap integral $s=0.129$. \cite{saito}  Fig.~\ref{1}(b) shows the EBS from the $30\times30$ supercell calculation. With the supercell EBS and corresponding wavefunctions, we unfold the energy bands to the FBZ of the primitive, and the results are shown in Fig.~\ref{1}(a) by red circles. Evidently, without symmetry breaking, the EBS unfolding of the supercell calculations indeed reproduces exactly the same EBS as that from the primitive cell calculation, indicating the validity of our unfolding approach.

\subsection{Unfolding bands of monolayer FeSe on SrTiO$_3$ substrate for \textit{ab initio} calculations}

\begin{figure}[bc]
  \centering
  \includegraphics[width =0.5\textwidth]{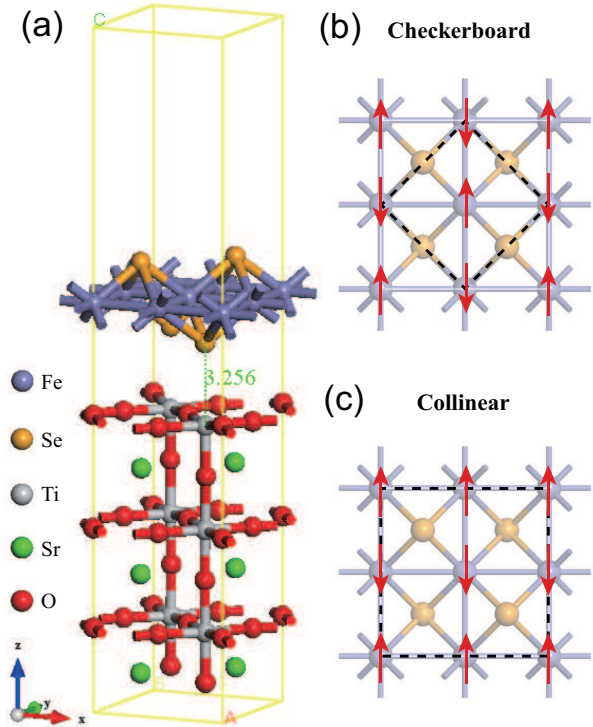}\\
  \caption{Atomic structure (a) and spin configurations of (b) checkerboard and (c) collinear AFM orders of Fe atoms for monolayer FeSe on TiO$_2$ terminated SrTiO$_3$ (001) surface. Spin-up and spin-down are marked by upward and downward red arrows, respectively. {The dashed-line squares in (c) and (d) denote $1\times 1$ primitive cell and $\sqrt{2}\times\sqrt{2}$ supercell respectively.}
}\label{2}
\end{figure}

Now we turn to a realistic but more complicated system. Recently, high-temperature superconductivity ($T_c\mathtt{\sim}65$ K) in monolayer FeSe grown on SrTiO$_3$ substrate (FeSe/STO) by molecular beam epitaxy was reported. \cite{QYWang} This simplest iron-based superconductor shows quite different Fermi surface topology from other iron-based superconductors, \cite{NComm,SLHe,DLFeng} presumably implying a different mechanism. Moreover, there are several intriguing issues that merit further studies. One is that the magnetic order of that system is still controversial until now. Recent ARPES experiment suggested that the ground state of FeSe/STO is spin density wave state, \cite{DLFeng} similar to its bulk conterpart. \cite{ZYLu} By using first-principles calculations, Liu \textit{et al.} \cite{KaiLiu} illustrated that the spin configuration of ground state FeSe/STO is collinear AFM state. The first-principles calculations by Bazhirov and Cohen \cite{Cohen} for the FeSe monolayer film without STO substrate, however, showed that the experimentally observed Fermi surface is best described by the checkerboard AFM spin configuration. And Zheng \textit{et al.} \cite{Fawei} further demonstrated the effects of charge doping and electric field on the Fermi surface of a monolayer checkerboard AFM FeSe on SrTiO$_3$.

\begin{figure}[bc]
  \centering
  \includegraphics[width =0.5\textwidth]{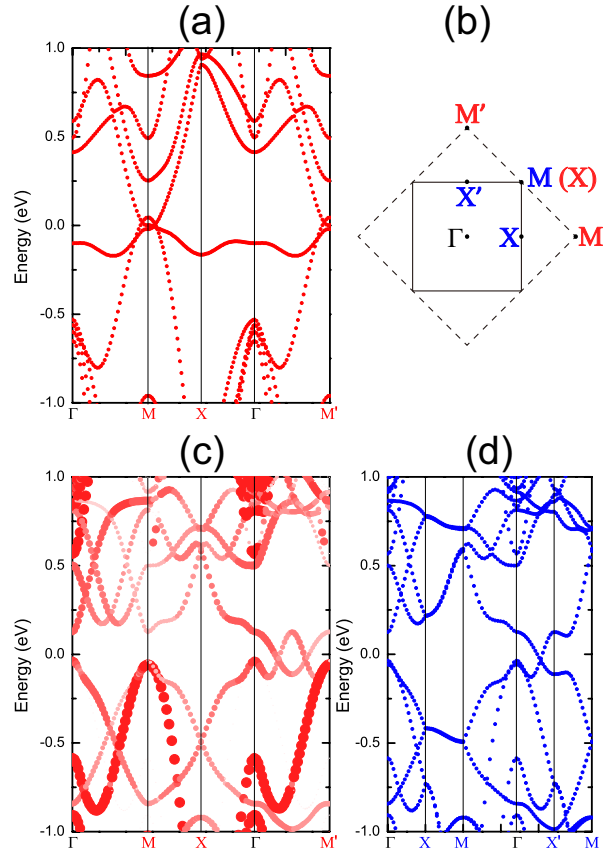}\\
  \caption{The spin-up EBS of monolayer FeSe/SrTiO$_3$ in different magnetic orders. (a) The spin-up EBS of FeSe/SrTiO$_3$ system in the checkerboard AFM state along high symmetry direction of the primitive-cell FBZ. (b) The FBZ of  primitive $1\times 1$ ($\sqrt{2}\times\sqrt{2}$) lattice with high symmetry $k$ points labeled by red (blue). (c) The unfolded and (d) folded spin-up EBSs of the system in the collinear AFM state. The shade and radius of dots in (c) represent the spectral weight of each eigenstate. The Fermi level is set to zero.
}\label{3}
\end{figure}

Actually, FeSe/STO systems in different magnetic states have different supercells and thus different FBZs. This makes it difficult to compare the calculated EBSs and Fermi surfaces in different FBZs with those from ARPES measurements which are in the FBZ of the primitive cell. \cite{oneFetwoFe}
Herein, by combining \textit{ab initio} calculation with our group-theory-based unfolding method, we studied FeSe/STO in the checkerboard (N\'{e}el) and collinear (striped) AFM spin configurations.

The calculations were carried out within the framework of DFT as implemented in the {\small QUANTUM ESPRESSO} code package. \cite{qe} A kinetic energy cutoff of 120 Ry was used for the plane-wave basis to achieve a balance between calculation efficiency and accuracy, and we adopted the generalized gradient approximation with Perdew-Burke-Ernzerhof exchange correlation functional. \cite{pbe} The norm-conseving pseudopotentials \cite{rrkj} were employed to describe the electron-ion interactions. The convergence thresholds of energy and force were set to $5.0\times 10^{-5}$ Ry and $5.0\times 10^{-4}$ Ry/Bohr for structure optimization. We constructed a six-layer SrTiO$_3$ (001) slab with monolayer FeSe on the TiO$_2$ termination, where the four bottom atomic layers are fixed at their bulk positions. To model checkerboard and collinear AFM configurations, we used $1\times1$ and $\sqrt{2}\times \sqrt{2}$ supercell with a vacuum layer of about 10 \AA. $9\times9\times1$ and $6\times6\times1$ Monkhorst-Pack $k$-point meshes \cite{mp} for Brillouin-Zone sampling were used for $1\times 1$ and $\sqrt{2}\times\sqrt{2}$ supercells respectively.

The optimized structure  of $\sqrt{2}\times\sqrt{2}$ FeSe/STO is shown in Fig.~\ref{2}(a). The lower Se (Fe) atoms are on the top of Ti (O) atoms with a average vertical distance of 3.26\AA~(4.56 \AA).
The checkerboard and collinear AFM spin configurations of Fe atoms are presented in Figs.~\ref{2}(b) and (c), respectively. The calculated magnetizations of checkerboard and collinear AFM states are, respectively, 2.77 and 3.02 $\mu_B$ per Fe atom, consistent with previous theoretical results. \cite{Cohen,Fawei}

The calculated spin-up EBSs are shown in Fig. 3. Note that spin-down EBS is similar to spin-up one, and thus is not shown for simplicity. As shown Fig. 3(a), FeSe/STO in the checkerboard AFM state has a Fermi surface pocket near the zone corners (i.e., $M$ and $M^\prime$ points), but without any indication of pockets around the zone center (i.e., $\Gamma$ point), which is consistent with recent experimental observations \cite{NComm}. Then, based on the spin-up EBS of the collinear AFM state with $\sqrt{2}\times\sqrt{2}$ supercell [see Fig. 3(d)], we get the unfolded EBS [see Fig. 3(c)] in the primitive FBS, where the shade and radius of dots represent the weight of each eigenvalues. Comparing the EBS of the checkerboard AFM state with the unfolded EBS of the collinear AFM state, we can see that hole-like bands appear in both zone corner and center in Fig. 3(c). Moreover, the unfolded bands cross the Fermi level along $\Gamma$$M^\prime$, indicating extra Fermi surface pockets located around $\Gamma$ towards $M^\prime$. Our analysis suggests that the checkerboard instead of collinear AFM state can yield the EBS compatible with recent experimental measurements.\cite{NComm,SLHe,DLFeng} This reveals that the ground state spin configuration of monolayer FeSe on TiO$_2$ terminated SrTiO$_3$ (001) surface may have checkerboard AFM order, which would be helpful to understand the superconductivity mechanism in low-dimensional Fe-based superconductors.

\section{Conclusion}
To summarize, we present a method to unfold the EBS from supercell calculations to the FBZ of the primitive lattice. Directly derived from the translation group symmetry, this method gives an uniform description of unfolding approaches based on various basis sets (such as plane waves, atomic orbitals and Wannier functions), which makes it easy to implement in both tight-binding model and existing ab initio code packages using different basis sets. Based on this method, we can easily compare unfolded EBSs of different supercell calculations and connect the theoretical results with ARPES measurements. As an example, we apply this method to investigate magnetic order in monolayer FeSe on SrTiO$_3$ (001) surface, finding that the checkerboard AFM state rather than the collinear AFM state has the EBS compatible with recent ARPES data. Our method can be further employed in study of defects, disorder, interfacial reconstructions and other systems that require the use of supercells.

\ack
H.H. and F.Z. equally contributed to this work. We thank Yong Xu for valuable discussions. This work is supported by the Ministry of Science and Technology of China (Grants Nos. 2011CB921901 and 2011CB606405) and the National Natural Science Foundation of China (Grants Nos. 11004013 and 11074139).

\appendix

\section{unfolding formula on the choice of plane wave as basis functions}
\label{sec:appendixA}
{To prove the equivalence and generality of our method, we derive the unfolding formula on the plane wave basis and compare our results with previous work.
The periodic part of supercell Bloch function can be expressed as:
\begin{equation}
|u'_{\vec{K}}\rangle=\sum_{\vec{G}}C_{\vec{K}}(\vec{G})e^{i\vec{G}\cdot \vec{x}},
\label{EqA1}
\end{equation}
where $\vec{G}=\sum^D_i N_i \vec{B}_i$ are supercell reciprocal lattice vectors.
We construct the projection operator $P(\vec{\mathcal{K}}^\prime)$ using a set of plane wave functions with primitive cell period
\begin{equation}
|\vec{g}, \vec{\mathcal{K}}^\prime\rangle= e^{i(\vec{\mathcal{K}}^\prime+\vec{g})\cdot \vec{x}},
\end{equation}
where $\vec{g}=\sum^D_i n_i \vec{b}_i$ are primitive cell reciprocal lattice vectors.
The expectation value of $P(\vec{\mathcal{K}}^\prime)$ is
\begin{equation}
\langle u'_{\vec{K}}|P(\vec{\mathcal{K}}^\prime)|u'_{\vec{K}}\rangle
=\sum_{\vec{g}} \langle u'_{\vec{K}}|{\vec{g}}, \vec{\mathcal{K}}^\prime\rangle \langle {\vec{g}}, \vec{\mathcal{K}}^\prime|u'_{\vec{K}}\rangle
=\sum_{\vec{g}} |\langle {\vec{g}}, \vec{\mathcal{K}}^\prime|u'_{\vec{K}}\rangle|^2.
\end{equation}
We use (A.1) and (A.2) to get
\begin{equation}
\begin{array}{l}
\langle \vec{g}, \vec{\mathcal{K}}^\prime|u'_{\vec{K}}\rangle= \sum_{\vec{G}}\int d\vec{x}C_{\vec{K}}(\vec{G})e^{i\vec{G} \cdot \vec{x}} e^{-i(\vec{\mathcal{K}}^\prime+\vec{g})\cdot \vec{x}}\\
= \sum_{\vec{G}} C_{\vec{K}}(\vec{G}) \delta(\vec{G}-\vec{g}-\vec{\mathcal{K}}^\prime)\\
=C_{\vec{K}}(\vec{g}-\vec{\mathcal{K}}^\prime),
\end{array}
\end{equation}
thus
\begin{equation}
\langle u'_{\vec{K}}|P(\vec{\mathcal{K}}^\prime)|u'_{\vec{K}}\rangle=\sum_{\vec{g}}|C_{\vec{K}}(\vec{g}-\vec{\mathcal{K}}^\prime)|^2.
\end{equation}
(A.5) is exactly the same as the (15) of Ref.~\cite{zunger_prb}, implying that our method can derive the equivalent results to previous work explicitly if plane wave functions are adapted as basis. What is more, our method does not need any information about the primitive eigenfunctions, which simplifies the derivation procedure.}


\end{document}